\documentclass[12pt,preprint]{aastex}

\tighten
\newcommand\be{\begin{equation}}
\newcommand\ee{\end{equation}}

\shortauthors{Antia and Basu}
\shorttitle{Temporal variations of the solar rotation rate}
\begin{document}

\title{Temporal variations of solar rotation rate at high latitudes}
\author{H. M. Antia}
\affil{Tata Institute of Fundamental Research,
Homi Bhabha Road, Mumbai 400005, India}
\email{antia@tifr.res.in}
\and
\author{Sarbani Basu}
\affil{Astronomy Department, Yale University, P. O. Box 208101,
New Haven CT 06520-8101, U.S.A.}
\email{basu@astro.yale.edu}

\begin{abstract}
Frequency splitting coefficients from Global Oscillation Network
Group (GONG) and Michelson Doppler Imager (MDI) observations covering
the period 1995--2001 are used to
study temporal variations in the solar rotation rate at high
latitudes. The torsional
oscillation pattern in the Sun is known to
penetrate to a depth of about $0.1R_\odot$ with alternate bands of
faster and slower rotating plasma. At lower latitudes the bands
move towards equator with time. At higher latitudes, however, the bands
appear to move towards the poles. This is similar to the observed pole-ward movement 
of large scale magnetic fields at high latitudes.
This also supports theoretical results of pole-ward moving bands at high
latitudes in some mean field dynamo models.
The polar rotation rate is found
to decrease between 1995 and 1999 after which it has started increasing.

\end{abstract}

\section{INTRODUCTION}

The rotation rate of the solar interior can be determined by
inverting observed rotational splittings of solar oscillation frequencies
(Thompson et al.~1996; Schou et al.~1998).
With the accumulation of GONG and MDI data over the
last six years, it is now possible to study the temporal variation of the
rotation rate in the solar interior.
The rotation rate is known to show
temporal variations, with bands of faster and slower rotating regions
moving towards the equator with time (Schou 1999;
Howe et al.~2000; Antia \&
Basu 2000) similar to torsional oscillations observed at the
solar surface (Howard \& LaBonte 1980; LaBonte \& Howard 1982;
Snodgrass 1992).
This pattern is found to penetrate to a depth of
about $0.1R_\odot$.
Torsional oscillations are believed to arise from nonlinear
interactions between magnetic field and differential rotation.
As such, they should provide a constraint on theories of solar dynamo.
Covas et al.~(2000) considered an axisymmetric mean field dynamo model
to study temporal variations in rotation rate and magnetic field in
solar interior. They find
temporal variation in the rotation rate which shows pattern similar
to torsional oscillations at low latitudes, with bands of faster and
slower rotating regions moving towards the equator with time.
But at high latitudes they find that these bands
migrate pole-wards. As far as surface observations go, some  magnetic features are seen
migrating pole-wards at high latitude (Leroy \& Noens, 1983; Makarov \& Sivaraman 1989).
This pole-ward movement may be crucial for magnetic field reversal
during the solar cycle.
Thus it is of interest to check if the observed zonal flow
pattern also moves pole-wards at high latitudes.
Inverters thus far have largely ignored the 
variation of the  rotation rate near the polar regions, where the
measurement of the  rotation rate is likely to be less reliable. 
Preliminary helioseismic work by Howe et al.~(2001) and
Basu \& Antia~(2001) suggests that 
that there could be a pole-ward flow at high latitudes.
In this work, we
make an attempt to study the temporal variations of the rotation rate near
the poles. We first test the ability of inversion techniques to resolve the
rotation rate near the poles using artificial data.
We then apply these techniques to study the temporal
variations in the rotation rate using observed splitting coefficients.

\section{DATA AND TECHNIQUE USED}

We have used data sets from GONG and MDI for this investigation. These sets
consist of the mean frequency and the splitting coefficients
for each $(n,\ell)$ multiplet.
We use the GONG data for months 1-55, which cover the period
from 1995 May 7 to 2000 October 6. We use all available 54 data
sets each covering a period of 108 days with a spacing of 36 days
between consecutive data sets. The first data set covers only 36 days
(a GONG month being 36 days).
The MDI data (Schou 1999) consist of 23 non-overlapping
data sets each covering a period of
72 days, starting from 1996 May 1 and ending on 2001 April 4,
with some gap in between when the SOHO satellite was out of contact.

We use a 2-dimensional Regularized Least
Squares (RLS) inversion technique to infer the rotation rate in the
solar interior from each of the available data sets.
The details of inversion technique are described by
Antia et al.~(1998). Since in this work we are mainly interested
in the  rotation rate at high latitudes where the inversion technique
are somewhat uncertain, we first perform a series of tests using artificial
data sets to ascertain the reliability of the inversion procedure.
For this purpose, we use a model rotation profile and calculate the
splitting coefficients for the assumed rotation rate. Then we add
random errors with same standard deviation as the estimated errors
in observed data sets and perform inversion using the same
regularization as that used for the real data. 
We construct artificial data sets using test profiles of the form
\be
\Omega(r,\theta)=460-55\cos^2\theta-55\cos^4\theta+
A\exp(-({r-r_0\over w_r})^2)\exp(-({\cos^2\theta-c_0\over w_c})^2)
\label{testprof}
\ee
where $\Omega(r,\theta)$ is the rotation rate in nHz and
$A, r_0, w_r, c_0, w_c$ are suitably chosen constants.
In Fig.~1, the inverted
profiles are compared with the actual ones for a few of these sets using
errors in MDI data sets.
It is clear that inversions are able to reproduce
relatively sharp peaks at high latitudes too, although the exact
shape may not match the actual peak. The radial extent of the peak
is reasonably well reproduced by the inversions.
Thus the inversion results appear to be reliable even at high latitudes.

\section{RESULTS}

To identify the time varying component of rotation rate, we
take the time average of all the results obtained from the different 
GONG (or MDI) data sets, and then subtract
this mean from the rotation rate at each epoch to get the
residual. This residual, $\delta\Omega$, contains the time-varying
part of the rotation rate. Fig~2 shows the contours of constant residual as a
function of time and
latitude at a depth of $0.02R_\odot$ below the solar surface as
obtained from both GONG and MDI data. 
The MDI results have a gap during 1998-99
when no data were available due to problems with the SOHO spacecraft.
In order to facilitate comparison
of the inversion  results with surface observations we show the rotation
velocity $v_\phi=r\Omega\cos\theta$, where $\theta$ is the latitude.
Also shown in the figure
are the contours obtained for surface rotation velocity using
Doppler measurements from Mt.\ Wilson (Ulrich 2001).
For the figure with Doppler results, we show the north-south symmetric
part of the rotation velocity since that is what the inversion
results determine.
In addition, we bin the  Doppler data over the time intervals
covered by the GONG data for a better comparison.         
In all  these results we can see that bands of
faster and slower rotation move towards equator at low latitudes.
But beyond about $50^\circ$, the bands appear to move towards
the poles. Theoretical results of Covas et al.~(2000)
based on a  mean field dynamo model also show this feature.
The latitude at which the transition
from equator-ward to pole-ward movement takes place is also similar
in their models.
The pole-ward drift is clear in the GONG results which do not have
any gap. The surface rotation rate as measured using Doppler techniques
(Ulrich 2001) also shows similar features, though the pole-ward
migration of bands at high latitude is not as clear as in the
helioseismic results, probably due to difficulties of making Doppler measurements at 
high latitudes.  Helioseismic data represents an average rotation
rate over the period of observations and this averaging smoothes out
variations on short time scales in rotation rate, which have been
observed in both Doppler measurements at solar surface (Ulrich et al.
1988; Hathaway et al. 1996) and in results from local helioseismic
techniques (Basu \& Antia 2000). 
To test the robustness of the pattern we have tried many
different regularization parameters for inversion and all results show
these bands. The GONG and MDI pattern will not match exactly as the
temporal mean which is subtracted to calculate the residuals,
is taken over different time intervals.

In addition to having these bands, the rotation
rate in the polar region has been decreasing with time during the
period 1995--1999. To show this variation clearly we show in Fig.~3
the rotation rate residuals at constant latitude as a function of time at a 
depth of $0.02R_\odot$ below the
solar surface. It is clear that there is a good agreement
between the GONG and MDI results. At the  latitude
of $85^\circ$, which is the highest latitude for which we have tried
to calculate inversion results,
both GONG and MDI results  show a clear decrease
in the rotation rate residuals  from 1995 to 1999, after which the residuals 
start increasing.
The minimum in polar rotation rate is found to be in early 1999 in
GONG data and slightly later in MDI data. At a latitude of $75^\circ$
the pattern of variation is similar, but the amplitude of variation
is much lower. The amplitude decreases rapidly as we move away from
pole. It may be noted that in this figure we have plotted, the
rotation rate, $\Omega$ rather than the rotation velocity $v_\phi$.
The amplitude of $v_\phi$ variation does not increase as we approach
the pole.

To study the depth dependence of the changes  in the polar rotation
rate we show, in Fig.~4, the rotation rate residuals at a latitude 
of $85^\circ$ at different depths. At $r=0.98R_\odot$ the errors are small and
both GONG and MDI data show a clear time-variation with a  minimum in
early 1999 as discussed earlier. This panel also shows the
North-South symmetric component of surface
rotation rate as inferred from Mt.\ Wilson Doppler measurements
(Ulrich 2001). It is clear that although the Doppler measurements
have larger fluctuations, the basic trend is similar to the
helioseismic results and in particular, these measurements also
show the minimum at around the same period.
The results for deeper layers have larger errors,
but even so a similar variation is
seen at $r=0.95R_\odot$. At even deeper layers there is no clear temporal variation.
Thus it seems possible that the depth to which the temporal variation
extends is similar to $0.1R_\odot$ found for the zonal flow
pattern at low latitudes (Howe et al.~2000; Antia \& Basu 2000).

\section{CONCLUSIONS}

The solar rotation rate obtained from inversions of different sets of 
GONG and MDI data are used to study the time evolution of the 
rotation rate.  The rotation-rate residuals, obtained by
subtracting the time-averaged rotation rate from that at
each epoch, show the
well known pattern of temporal variation similar to the torsional
oscillations observed at the surface,  with bands of faster and
slower rotation moving towards the equator with time at low
latitudes.
At high latitudes it appears that the
bands  move towards the pole instead, the transition between equator-ward
and pole-ward movement appears to be around a latitude of $50^\circ$.
Observations of magnetic features also show such pole-ward movement 
at high latitudes (see Leroy \& Noens 1983; Makarov \& Sivaraman 1989; 
Erofeev \& Erofeeva 2000; Benevolenskaya et al.~2001). 
Theoretical results of Covas et al.~(2000) using a mean-field
dynamo model also show this feature.
Our inversion results therefore re-inforce
the link between zonal flows and the solar magnetic cycle. 
Earlier works too have shown connections between 
zonal flows and the solar activity cycle (Antia \& Basu 2000).

The rotation rate in the outer layers of the  polar regions
varies with time, reaching a minimum in early
1999 after which it has started increasing. The time of minimum
rotation rate at poles is distinctly before the maximum in solar
activity. Similarly, it appears that the maximum rotation
rate was achieved before the minimum in activity.
These changes appear to persist till a depth of about $0.1R_\odot$,
similar to the depth of penetration of zonal flow pattern at low
latitudes. The observed temporal variations in the rotation
rate should provide constraints on dynamo models.

\section*{ACKNOWLEDGEMENTS}

We would like to thank Roger Ulrich for the surface Doppler data 
and permission to use them.
This work  utilizes data obtained by the
Global Oscillation Network Group (GONG) and from
the Solar Oscillations
Investigation / Michelson Doppler Imager  on the Solar
and Heliospheric Observatory (SOHO). 
GONG is managed by the National Solar Observatory, which
is operated by AURA, Inc. under a cooperative agreement with the
National Science Foundation. The data were acquired by instruments
operated by the Big Bear Solar Observatory, High Altitude Observatory,
Learmonth Solar Observatory, Udaipur Solar Observatory, Instituto de
Astrof\'{\i}sico de Canarias, and Cerro Tololo Interamerican
Observatory.
This work also utilizes data from the Solar Oscillations
Investigation / Michelson Doppler Imager (SOI/MDI) on the Solar
and Heliospheric Observatory (SOHO).  SOHO is a project of
international cooperation between ESA and NASA.

\begin{figure}
\plotone{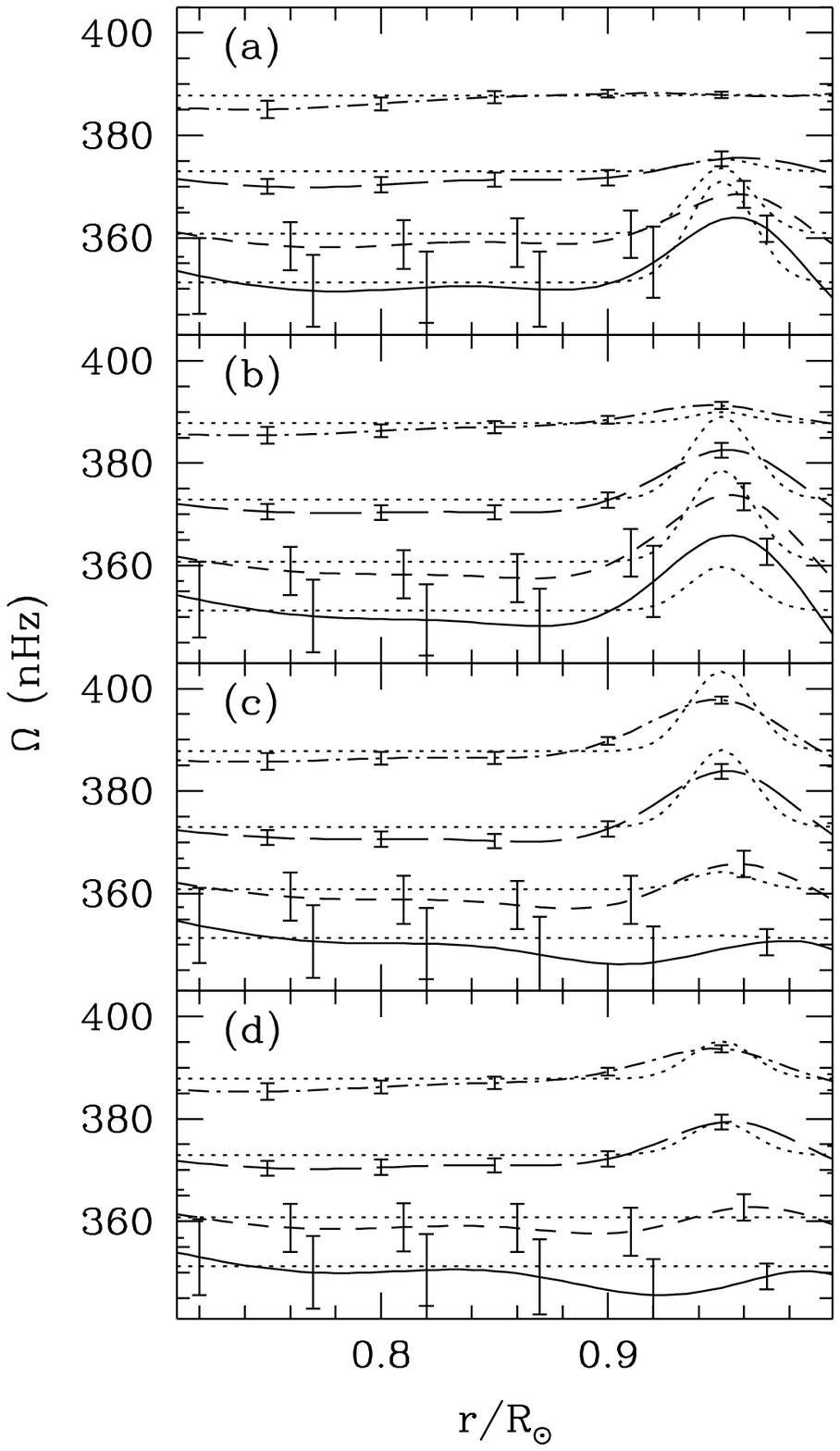}
\caption{Artificial data inversion results for several different
model rotation profiles. In each panel the dotted lines
show the true rotation rate, while other lines show the inverted
profiles at different latitudes. The continuous line is for
$85^\circ$, the short dashed line for $75^\circ$, the long dashed line for
$67.5^\circ$ and dot-dashed line for $60^\circ$ latitudes.
All the artificial profiles are for $r_0=0.95R_\odot$,
$w_r=0.02R_\odot$, $A=20$ nHz in Eq.~(1).
The panel (a) is for $c_0=1$, $w_c=0.1$,
panel (b) is for $c_0=0.9$, $w_c=0.1$,
panel (c) is for $c_0=0.8$, $w_c=0.1$, while
panel (d) is for $c_0=0.8$, $w_c=0.05$.}
\label{fig:art}
\end{figure}

\begin{figure} 
\plotone{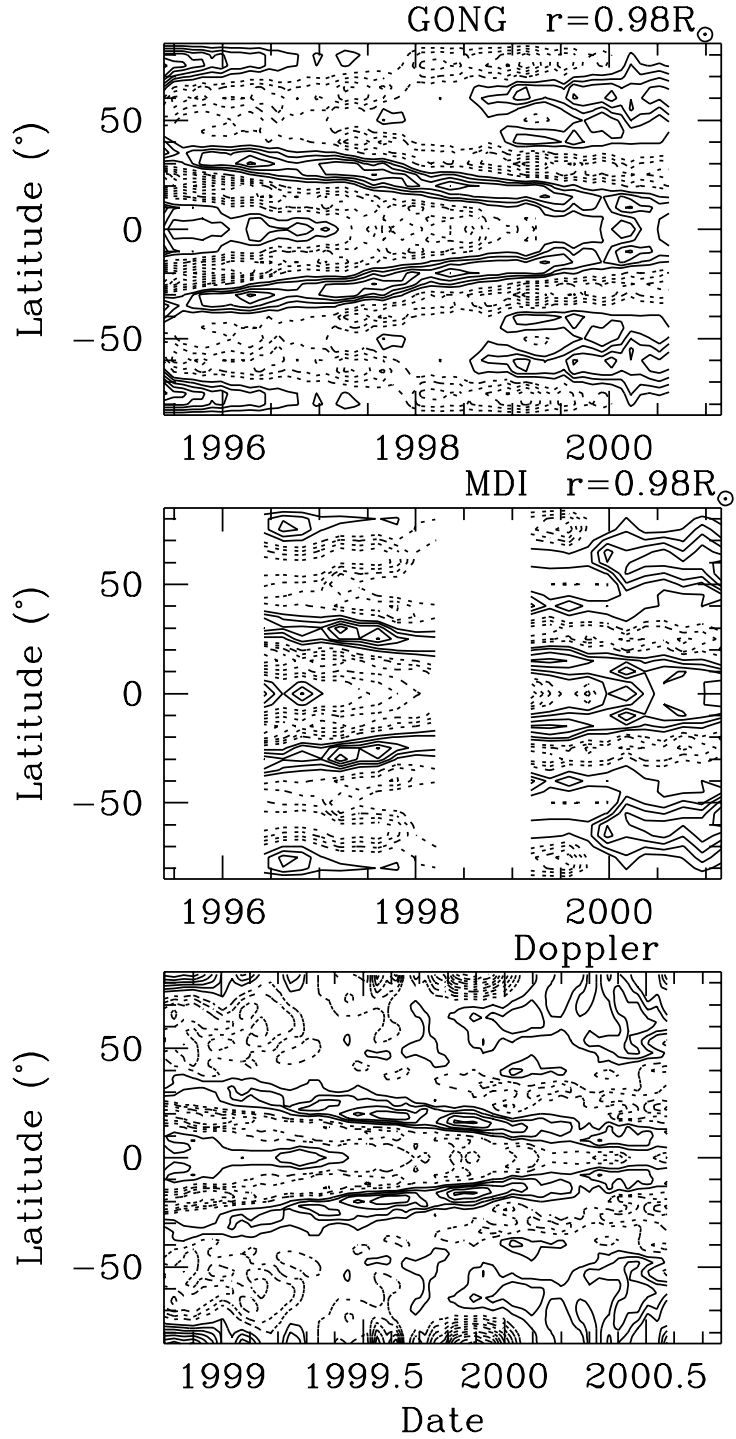}
\caption{Contour diagrams of constant rotation velocity residuals
at $r=0.98R_\odot$ obtained using 2D RLS inversion of
GONG and MDI data. Also shown are the residuals for
surface Doppler observation of Ulrich (2001).
The continuous contours are for positive 
$\delta v_\phi$, while dotted contours denote negative values.
The contours are drawn at interval of 1 m/s.}
\end{figure}

\begin{figure}
\plottwo{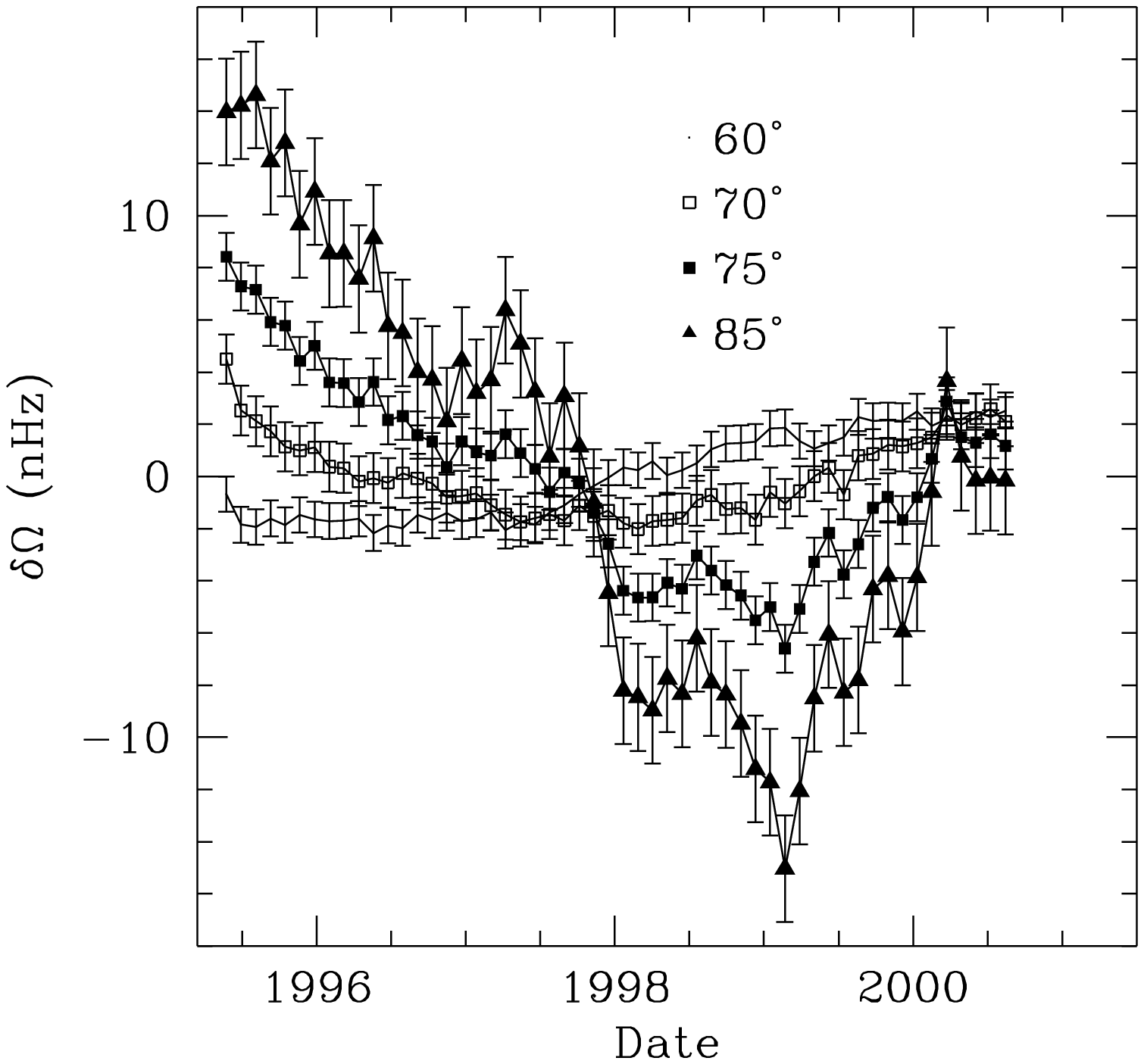}{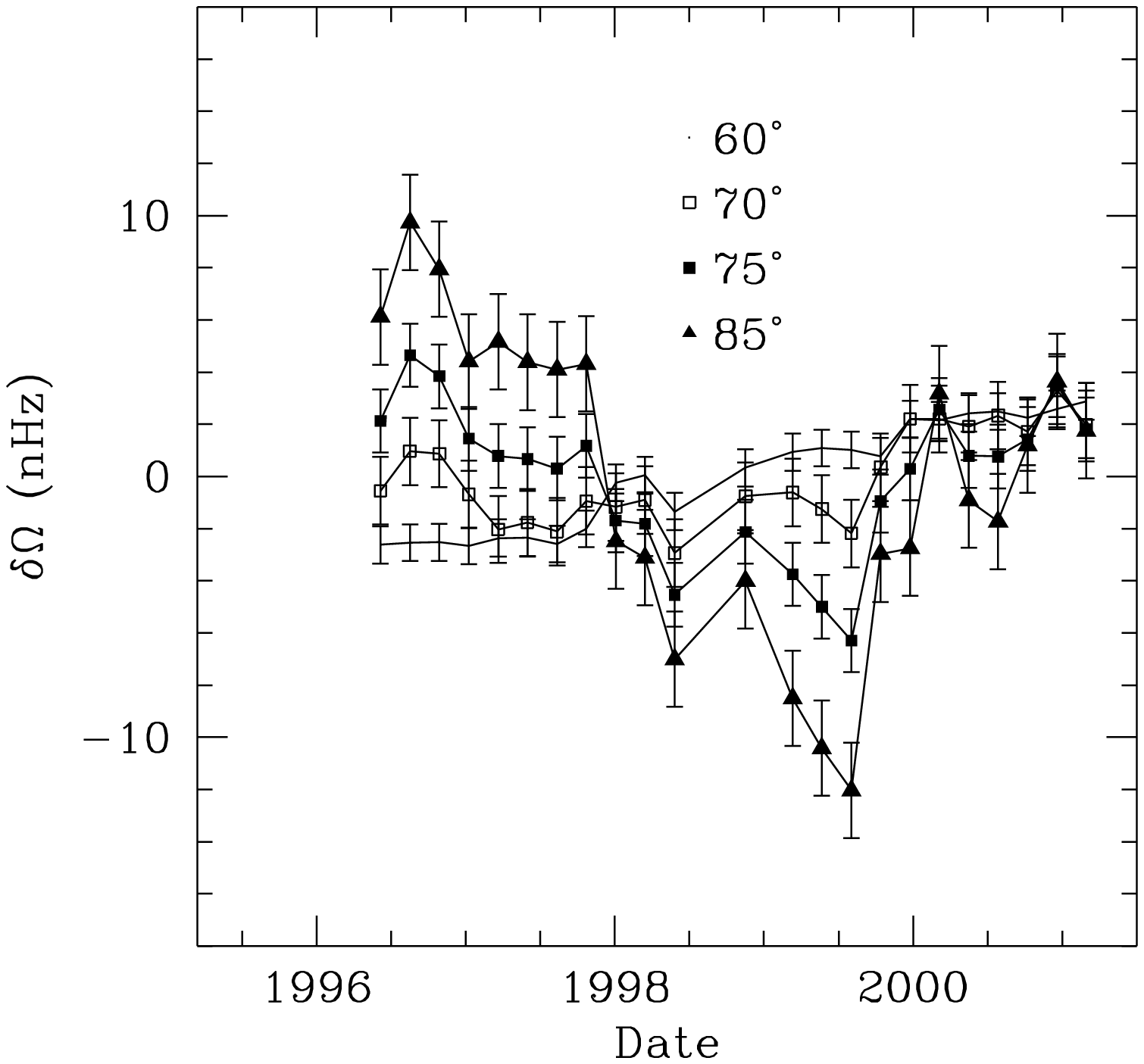}
\caption{The rotation-rate residuals at 
$r=0.98R_\odot$ plotted as a function of
time for different latitudes.
The latitudes are marked in the figure.
The results were obtained using 2d RLS inversion of GONG (left panel)
and MDI (right panel) data.}
\end{figure}

\begin{figure}
\plotone{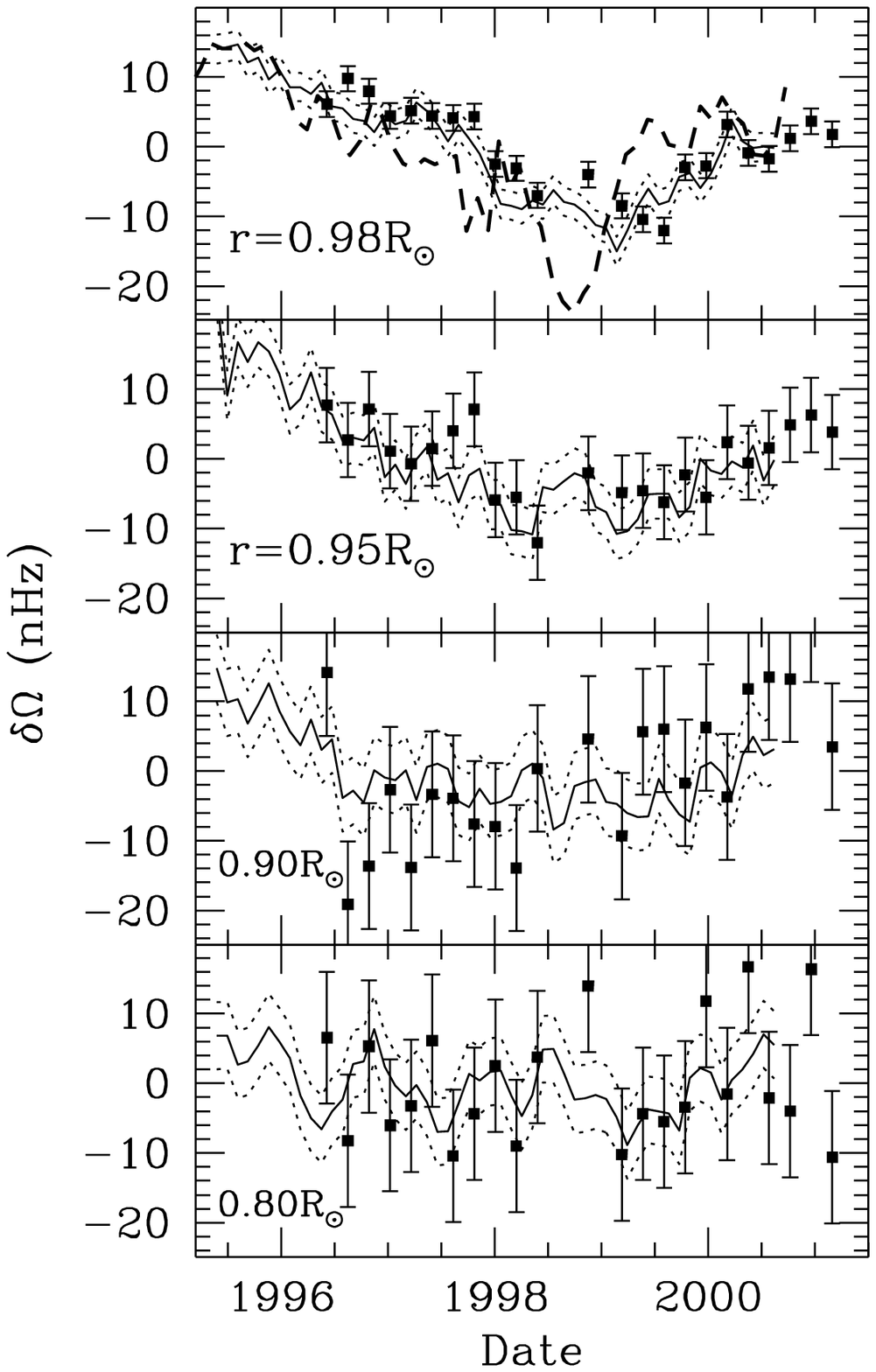}
\caption{The rotation-rate residuals  at the latitude of
$85^0$ plotted
as a function of time at a few selected radii as marked in each panel.
The results were obtained using 2d RLS inversion of GONG and MDI data.
The continuous lines show the GONG results with dotted lines showing
the $1\sigma$ error estimates, while the filled squares with error-bars
show the MDI results. The thick dashed line in the topmost panel
shows the results from Doppler measurement at the solar surface.}
\end{figure}


\begin{thebibliography}{}

\bibitem[\protect\astroncite{Antia}{2000}]{ant00}
Antia, H. M., \& Basu, S. 2000, ApJ, 541, 442

\bibitem[\protect\astroncite{Antia}{1998}]{ant98}
Antia, H. M., Basu, S., \& Chitre, S. M. 1998, MNRAS, 298, 543

\bibitem[\protect\astroncite{Basu}{2000}]{ba00}
Basu, S., \& Antia, H. M. 2000, J. Astrophys.\ Astron., 21, 353 

\bibitem[2001]{ba01} Basu, S., \& Antia, H. M.
2001, in Helio- and Asteroseismology at the Dawn of the Millennium,
ed. A. Wilson, ESA SP-464, p. 179

\bibitem[2001]{ben01}
Benevolenskaya, E. E., Kosovichev, A. G., \& Scherrer, P. H.
2001, ApJ, 554, L107

\bibitem[\protect\astroncite{Covas}{2000}]{cov00}
Covas, E., Tavakol, R., Moss, D., \& Tworkowski, A. 2000, A\&A, 360, L21

\bibitem[2000]{ero00}
Erofeev, D. V., \& Erofeeva, A. V. 2000, Sol.\ Phys., 191, 281

\bibitem[1996]{hat96}
Hathaway, D. H. et al.~1996, Sci., 272, 1306

\bibitem[1980]{how80} Howard, R., \& LaBonte, B. J. 1980, ApJ, 239, L33


\bibitem[\protect\astroncite{Howe}{2000a}]{how00}
Howe, R., Christensen-Dalsgaard, J., Hill, F., Komm, R. W.,
Larsen, R. M., Schou, J., Thompson, M. J. \& Toomre, J. 2000, ApJ, 533, L163

\bibitem[2001]{howe0} Howe, R., Christensen-Dalsgaard, J., Hill, F., 
Komm, R. W., Larsen, R. M., Schou, J.,  Thompson, M. J., \& Toomre, J. 
2001, in Helio- and Asteroseismology at the Dawn of the Millennium,
ed. A. Wilson, ESA SP-464, p. 19

\bibitem[1982]{lab82}
LaBonte, B. J., \& Howard, R. 1982, Sol.\ Phys., 75, 161

\bibitem[1983]{le83}
Leroy, J. -L., \& Noens, J. -C. 1983, A\&A, 120, L1

\bibitem[1989]{mak89}
Makarov, V. I., \& Sivaraman, K. R. 1989, Sol.\ Phys., 123, 367


\bibitem[\protect\astroncite{schou}{1999}]{sch99}
Schou, J. 1999, ApJ, 523, L181


\bibitem[\protect\astroncite{Schou}{1996}]{sch98}
Schou, J. et al.\ 1998, ApJ, 505, 390

\bibitem[1992]{sno92} Snodgrass, H. B. 1992, in The Solar Cycle,
proc. NSO 12th Summer Workshop, ASPS 27, p205


\bibitem[\protect\astroncite{Thomp}{1996}]{tho96}
Thompson, M. J. et al.\ 1996, Sci., 272, 1300

\bibitem[1988]{ulr88}
Ulrich, R. K. et al. 1988, Sol.\ Phys., 192, 437

\bibitem[2001]{ulr01}
Ulrich, R. K. 2001, ApJ, submitted

\end{thebibliography}
\end{document}